# Real-Field Hong–Ou–Mandel Interference of Indistinguishable Coherent Photons via Long Optical Injection-Locking over 50 km Fiber

Seoyeon Yang[1], Danbi Kim[1], Hansol Jeong[1], and Han Seb Moon[1,2,3*]

[1]*Department of Physics, Pusan National University, Geumjeong-Gu, Busan 46241, Korea*
[2]*Quantum Sensors Research Center, Pusan National University, Geumjeong-Gu, Busan 46241, Korea*
[3]*Quantum Science Technology Center, Pusan National University, Geumjeong-Gu, Busan 46241, Korea*

*\*hsmoon@pusan.ac.kr*

Measurement-device-independent quantum key distribution (MDI-QKD) has garnered significant attention for its potential to enable security-loophole-free quantum communication. Successful MDI-QKD protocols rely on performing a two-photon Bell-state measurement at an intermediate node, with a high-visibility Hong–Ou–Mandel (HOM) interference pattern between two independent coherent photons being crucial. In this study, we present a novel approach for developing indistinguishable coherent photon sources over 50 km of optical fiber in a real-world setting. We introduce the long optical injection-locking (long-OIL) technique, which enables frequency locking between two long-distance coherent photons beyond the coherence length of the master laser. Using the long-OIL technique, we achieved time-resolved HOM interference with a visibility of 48(2)%, approaching the theoretical 50% limit for two independent continuous-wave coherent photons. Our results demonstrate that the long-OIL platform is a promising solution for MDI-QKD with repeaterless secret key capacity.

**Introduction**

Quantum key distribution (QKD) enables two distant parties (Alice and Bob) to distribute secure keys [1–3]. Although commercial QKD systems are now available, QKD still exhibits numerous practical security vulnerabilities due to the use of imperfect devices [4–7]. Owing to this practical limitation, most commercial QKD systems utilize a weakly coherent photon source instead of a single-photon source or an entangled-photon source. The multiphoton problem associated with coherent photon sources can be mitigated using the decoy-state method [8, 9]. However, many attacks target the most vulnerable detection aspect of any QKD device [10, 11]. Concurrently, all security loopholes on the detection side are addressed by measurement-device-independent QKD (MDI-QKD) [12–16] protocols through the implementation of a two-photon Bell-state measurement at the intermediate node. Furthermore, various twin-field QKD (TF-QKD) protocols, based on coherent states for the utilization of remotely prepared optical fields with similar phases, surpass the repeaterless secret key capacity over long distances, and akin to MDI-QKD, employ an intermediate node that can be entirely untrusted [17–24].

Generally, the spectral, spatial, and polarization modes of coherent photons at the output beam splitter in a two-photon Bell-state measurement setup should be indistinguishable and superposed to realize high-visibility Hong–Ou–Mandel (HOM) interference [25–33]. Therefore, preparing two indistinguishable sets of distant coherent photons for the practical implementation of quantum communication systems using the MDI-QKD protocol [21–24] is essential. Here, an optical frequency difference between two completely independent lasers must be maintained within the laser's spectral linewidths to ensure spectral indistinguishability. Recently, frequency-stabilization techniques utilizing high-resolution spectroscopy of Rb atoms have been applied to observe HOM interference fringes with a visibility of 46%; this has been achieved using two independent sets of coherent photons at a distance of 600 m frequency-locked to the atomic transition between the hyperfine states [31]. Although this method guarantees the absolute preparation of many fully independent coherent photons regardless of the separation distance between the light sources, it is limited to the operating optical frequency conditional to the resonance frequency of atomic transitions. Meanwhile, indistinguishable coherent photon sources (ICPSs) based on frequency-stabilization techniques have been experimentally demonstrated at near-infrared wavelengths. Furthermore, the telecom-band quantum coherent photons are essential for realizing long-distance quantum communications and quantum networks.

However, the optical injection-locking (OIL) method is the most sophisticated technique for phase locking between two independent lasers without electronic control [34–39]. OIL has already been used in various applications such as optoelectronic millimeter-wave synthesis, optical filters for wavelength selection in dense-wavelength-division multiplexing, and component selection of optical frequency comb [38]. In the field of quantum communication, OIL has been applied to generate encoded bit states for MDI-QKD [23, 39]. For precise control of the phase between pulses and increasing laser modulation bandwidth, the phase-encoding laser is optically injected into a secondary laser through a circulator [23]. In a recent work, OIL was used to demonstrate a multirate, multiprotocol QKD transmitter [40]. On the other hand, the OIL has also been used in the MDI-QKD hacking strategy based on the trusted-source assumption [41].

In this work, we propose a novel approach for ICPSs based on a long-OIL system with a long-distance optical path greater than the coherence length of the master laser. We term the proposed method "long-OIL" to discriminate from the typically known OIL for phase locking between two lasers via the stimulated process of a slave due to a master laser. For the first time, we report the successful experimental demonstration of an ICPS based on a long-OIL system with an optical fiber tens of kilometers in length in the real field. In the setup, Alice's and Bob's distributed feedback (DFB) lasers act as master and slave lasers, respectively. The optical fiber is several thousand times longer than the coherence length of the DFB lasers. To confirm the indistinguishability of the two frequency-locked photon sources by applying the long-OIL system, we experimentally demonstrate time-resolved HOM interference between two ICPSs based on the long-OIL system connected to a 50 km-long optical fiber in the real field. We measure and compare the HOM interference fringes with two coherent photons with and without long-OIL using time-resolved two-photon detection.

**Experimental scheme for ICPSs based on long-OIL**
Figure 1 illustrates the experimental scheme for the realization of ICPSs (Alice and Bob) based on the long-OIL and time-resolved HOM interference setups (Charlie) utilizing two separate continuous wave (CW)-mode coherent photon sources. This experimental configuration, excluding the encoders for QKD, is similar to those of the MDI-QKD and TF-QKD protocols through the implementation of two-photon measurements at the intermediate node [23]. Alice and Bob, two spatially separated DFB lasers with specified optical frequencies and polarization modes, are required to be indistinguishable. In our experiment, two DFB lasers with a telecom wavelength of 1529 nm were independently operated and spatially separated. The Bob and Charlie parties were connected via an optical fiber with a length of 50 km in the real field (Korea Research Environment Open NETwork: KREONET).

We implemented the ICPSs of Alice and Bob using a long-OIL channel comprising a 25 km-long optical fiber in the experimental setup shown in Fig. 1. Alice's DFB laser integrated with an optical isolator as the master laser is optically injected into the Bob's DFB laser without an optical isolator as the slave laser. In the experiment, the spectral linewidth of the DFB lasers was measured to be approximately 5 MHz, corresponding to a coherence length of 60 m. Therefore, because the optical path length of the long-OIL channel between Alice and Bob is approximately 2000 times greater than the coherence length of the DFB lasers, the coherent photons of Alice and Bob are indistinguishable but not phase-locked.

In the setup, Alice's master laser is optically injected into slave laser of Bob via the long-OIL channel. Here, the circulator is used to optically inject Alice's laser and transmit Bob's coherent photons to Charlie.

Generally, the locking bandwidth of a typical OIL depends on the polarization, power, and mode matching between the master and slave lasers. In our system, which is based on single-mode optical fiber (SMF) components, spatial-mode matching is nearly complete. To optimize the long-OIL setup, we controlled the power and polarization of the injected laser using a variable optical attenuator (VOA) and a fiber polarization controller (FPC). To investigate the bandwidth of the long-OIL system as a function of the injection power, we measured the beat signal between Alice and Bob's lasers using a high sensitivity PIN amplified detector and spectrum analyzer before and after injection locking. The power of Alice's laser injected into the Bob section was measured as approximately 4 μW before the circulator, and the power of Bob's laser was measured as approximately 10 mW after the circulator. Within the locking-bandwidth range, the center frequency of Bob's slave laser is obeyed to that of Alice's master laser. Although Alice's initial polarizationis changed to an unknown polarization state, the injected polarization can be adjusted to the maximum locking bandwidth using the FPC.

Meanwhile, we note that the Charlie party is composed of the two-photon HOM interference between the coherent photons of Alice and Bob and the externally connected 50 km-long optical fiber network. In our experiment, the two-photon sources locked by the long-OIL setup afforded the advantage of indistinguishability without any electronic feedback. The two ICPSs were subsequently directed onto Charlie's 50:50 nonpolarizing beam splitter (BS). The polarizations of the two input photons were regulated by the FPCs positioned at the two SMF input ports. After their passage through the BS, the superposed output photons were detected using two SMF-coupled superconducting nanowire single-photon detectors (SNSPDs). The output signals from the two SNSPDs were transmitted to a time-correlated single-photon counter (TCSPC) for twofold coincidence counting and subsequent time-resolved measurement of the HOM interference of the two independent ICPSs [31, 32].

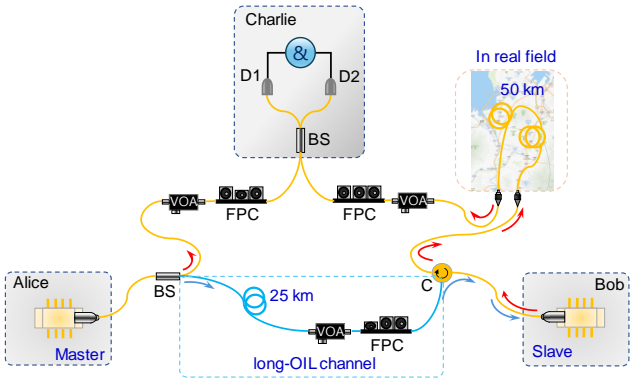

**Figure 1. Experimental configuration for ICPSs based on long-OIL.** ICPSs (Alice and Bob) based on long-OIL and time-resolved HOM interference (Charlie) with two separated CW-mode coherent photon sources with a 50 km-long optical fiber in a real field (BS, beam splitter; D, superconducting nanowire single-photon detector; C, circulator; VOA, variable optical attenuator; FPC, fiber polarization controller).

**Spectral properties of long-OIL**

Many previous studies and applications using OIL have been based on phase locking between the master and slave lasers because the path length for optical injection lies within the coherence length of the laser used. However, the scenario of our proposed long-OIL is assumed to involve a significantly greater distance between the two lasers. In this experiment, the phase relationship between Alice's coherent photons that depart from Alice and arrive at Bob's end may be instantaneously random. Therefore, the long-OIL scenario is significantly different from that of conventional OIL with regard to coherence or phase locking. The long-OIL setup raises an important question regarding its role and function in view of the stimulated coherence perspective involving the indistinguishability of photons.

To address this question, we measured the beat signal between Bob's slave laser upon applying long-OIL and Alice's master laser after shifting its frequency by 80 MHz using an acousto-optic modulator (AOM). Figure 2 shows the beat signal measured using an RF spectrum analyzer; the signal's spectral width was measured to be ~4.5 MHz. The measured center frequency of the beat signal is 80 MHz, corresponding to the driving frequency of the AOM, which means that Bob's optical frequency is locked to that of Alice. However, the observed spectral width indicates relative phase noise between Alice and Bob, which corresponds to the spectral linewidth of the DFB lasers used. Therefore, Bob's slave laser is not phase-locked to Alice's master laser. From the results shown in Fig. 2, we can conclude that long-OIL can contribute to the frequency locking of long-distance-separated coherent photon sources via an incoherently stimulated process. However, the phase of Bob's coherent photons is independent of that of Alice's.

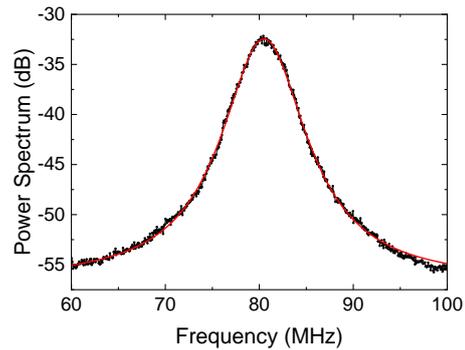

**Figure 2. Frequency-locking via long-OIL.** Beat signal between the frequency-locked Bob slave laser connected via long-OIL with 25 km-long optical fiber and the Alice master laser shifted by an 80 MHz AOM.

**Locking bandwidth of long-OIL**

In our long-OIL setup based on a fiber optic system, the spatial mode is well matched because of the use of SMF components, and the polarization mode for locking-bandwidth optimization is adjusted using polarization controllers. Figure 3(a) shows the optical analyzer signal of Bob's slave laser before (blue curve) and after (red curve) applying long-OIL using a Fabry–Pérot cavity with a free-spectral range of 1.5 GHz. We find that Bob's optical frequency follows that of Alice after applying long-OIL (red curve). In this case, the optical frequency difference was measured as $\omega_{diff} = 267(35)$ MHz between the free-running Alice and Bob lasers before applying long-OIL by comparing the optical frequencies of the blue and red curves under the condition of an injected power of 4 μW.

In this study, the locking bandwidth is defined as the maximum optical frequency difference between the free-running Alice and Bob lasers, which allows complete frequency locking owing to the long-OIL setup. The locking bandwidth changes as a function of the injected power of Alice's master laser passing through the long-OIL channel, as shown in Fig. 3(b). At an injection power of 12 μW, the measured locking bandwidth was approximately 760 MHz. Considering the optical frequency drift (less than 100 MHz in an hour) of the free-running Alice and Bob lasers, we believe that the locking bandwidth of our system is sufficient for hour-long operation. Although the locking bandwidth does not increase linearly as a function of the injection power, it does increase as the injection power increases.

$$P_{coin}(\tau) = 1 - V\Gamma_{AB}(\tau)\cos(\omega_{diff}\tau), \quad (1)$$

where V denotes the visibility of the HOM fringe, $\Gamma_{AB}(\tau)$ the mutual coherence function between the two input coherent photons, and $\omega_{diff}$ the optical frequency difference between Alice and Bob. If frequency locking via long-OIL implies $\omega_{diff} = 0$, $\Gamma_{AB}(\tau)$ determines the shape and width of the HOM fringe. In our experiment, the CW-mode coherent photons obtained using an SMF and its components are free in the temporal and spatial modes, and the polarization mode is controlled by the FPCs. Therefore, the photons from Alice and the frequency-locked Bob sources via the long-OIL channel are indistinguishable in the spectral, temporal, spatial, and polarization modes of the coherent photons.

Figures 4(a) and 4(b) demonstrate, for the first time, the successful measurement of the drastic change in the time-resolved HOM interference fringes before and after long-OIL implementation, obtained on connecting Bob and Charlie via an external 50 km-long optical fiber. The horizontal axis represents the detection time delay between the two SNSPDs, where the time bin is set to 0.5 ns. The coincidence counting rates accumulated over 60 s are normalized to 1 to ensure a significantly longer time delay than the mutual coherence time. Figure 4(a) shows the HOM-type beating fringe including two-photon beating under the condition of $\omega_{diff}$ = 153 MHz and Alice and Bob's free-running in the absence of long OIL. In addition, without long-OIL, the frequency difference changes over time, causing the shape of the fringe to change. However, by implementing the ICPSs of Alice and Bob using a 25 km-long optical fiber long-OIL channel, the beat HOM fringe shown in Fig. 4(a) dramatically changes to the fringe in Fig. 4(b) with a visibility as high as 48%, close to the visibility of the 50% theoretical maximum. From the results of high-visibility HOM interference, we confirmed that our long-OIL method is useful for the generation of ICPSs. Furthermore, because the variation in the frequency difference with time lies within the locking bandwidth, the HOM fringe is stably maintained.

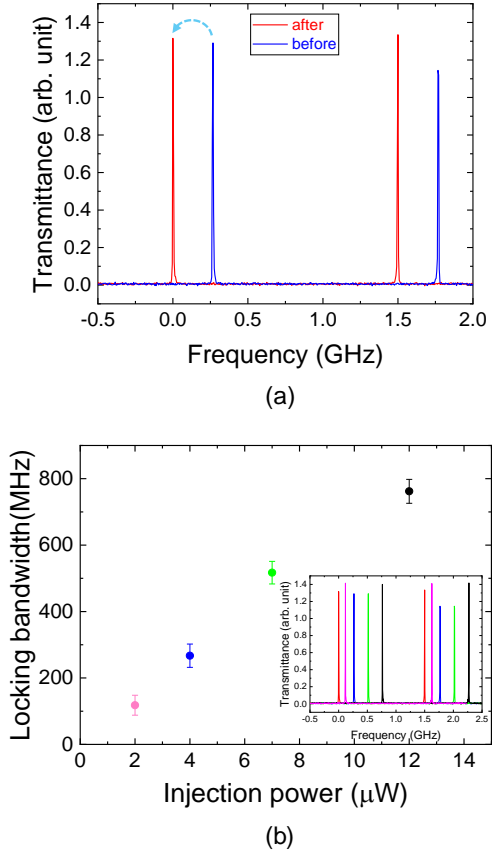

**Figure 3.** Frequency-locking bandwidth of long-OIL. (a) Under the condition of an injected power of 4 μW and optical frequency difference $\omega_{diff}$ = 267(35) MHz between Alice and Bob before (blue curve) and after (red curve) applying the long-OIL frequency, with the cavity transmittance signal of the frequency-locked Bob laser corresponding to the optical frequency of Alice's master laser (red curve). (b) As a function of the power injected into the Bob laser, with the frequency-locking bandwidth from the measured optical cavity modes of Bob's slave laser before applying long-OIL frequency: the inset graph shows the cavity transmittance signals (the data-point colors correspond to the transmittance signal colors).

**Time-resolved HOM interference with ICPSs based on long-OIL**

To confirm the indistinguishability of both the frequency-locked coherent photons obtained via long-OIL, we demonstrated time-resolved HOM interference with two ICPSs with a long-OIL channel for optical synchronization. When two CW-mode coherent photons of Alice and Bob meet in the two input ports of the BS in the Charlie section of the experimental setup, the coincidence counting probability as a function of the time delay (τ) between the start and stop of the TCSPC can be expressed as [42, 43]

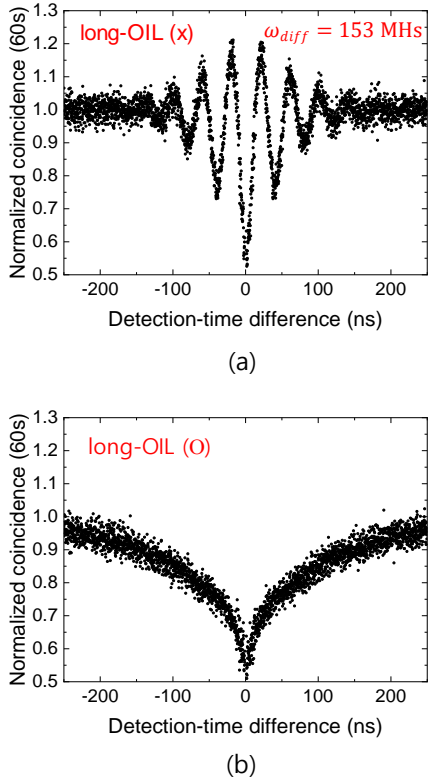

**Figure 4. Time-resolved HOM interference with ICPSs based on long-OIL.** (a) HOM-type two-photon beating fringes under the condition of a frequency difference of $\omega_{diff}$ =153 MHz between Alice and Bob's optical frequencies without long-OIL. (b) HOM interference fringe with ICPSs after frequency-locking upon applying long-OIL.

**Conclusions**

In this study, we experimentally demonstrated the proposed "long-OIL" method with a long-distance optical path between Alice's master laser and Bob's slave laser that is greater than the coherence length of Alice's photons. We successfully demonstrated high-visibility HOM interference between two stable ICPSs based on a long-OIL setup connected to a 50 km-long optical fiber in a real-world setting. Because the proposed long-OIL assumes a significantly long distance between both lasers, we confirmed that the long-OIL platform contributes to the frequency locking of long-distance-separated coherent photon sources via an incoherently stimulated process. Furthermore, we report not only a substantial change in HOM interference fringes before and after long-OIL implementation, but also a stable and high-visibility HOM signal due to the broad locking bandwidth of the long-OIL. If the length of the long-OIL channel between Alice and Bob is considerably large, amplifying the power of Alice's master laser to overcome the optical fiber loss is possible. Consequently, multiple ICPSs based on the long-OIL method may be beneficial for realizing long-distance quantum communication and quantum networks.

**Funding.**
This study was supported by Institute for Information and Communications Technology Planning and Evaluation (IITP) (Nos. RS-2024-00396999, IITP-2025-2020-0-01606, and IITP-2022-0-01029), National Research Foundation of Korea (NRF) grant funded by the Korean Government (MSIT) (No. RS-2023-00283146), and the Regional Innovation Strategy (RIS) through the NRF, funded by the Ministry of Education (MOE) (2023RIS-007).

**Acknowledgement**.
This paper was supported by KISTI/KREONET.

**Disclosures.**
The authors declare no conflicts of interest.

**Data availability.** Data underlying the results presented in this paper may be available from the corresponding author upon reasonable request.